\documentclass[12pt]{iopart}
\usepackage{graphicx}
\usepackage{bm}
\usepackage{subfigure}

\begin{document}

\title{Boundary conditions in random sequential adsorption}

\author{Micha\l{} Cie\'sla$^1$, Robert M. Ziff$^2$}
\address{$^1$ M.\ Smoluchowski Institute of Physics, Department of Statistical Physics, Jagiellonian University, \L{}ojasiewicza 11, 30-348 Krak\'ow, Poland.}
\address{$^2$ Center for the Study of Complex Systems and Department of Chemical Engineering, University of Michigan, Ann Arbor MI 48109-2136 USA.}%
\ead{michal.ciesla@uj.edu.pl, rziff@umich.edu}

\date{\today}

\begin{abstract}
The influence of different boundary conditions on the density of random packings of disks is studied. Packings are generated using the random sequential adsorption algorithm with three different types of boundary conditions: periodic, open, and wall. It is found that the finite size effects are smallest for periodic boundary conditions, as expected. On the other hand, in the case of open and wall boundaries it is possible to introduce an effective  packing size and a constant correction term to significantly improve the packing densities. 
\end{abstract}
\pacs{02.70.Tt, 05.10.Ln, 68.43.Fg}
\vspace{2pc} \noindent{\it Keywords}: random paskings, random sequential adsorption, periodic boundary conditions, finite size effecst
\maketitle

%
%
%
\section{\label{sec:introduction}Introduction}
Random sequential adsorption (RSA) refers to a numerical procedure popularized by Feder and others to model monolayers obtained in irreversible adsorption processes~\cite{feder1980}. It is based upon sequential iterations of the following steps:
\begin{itemize}
\item[--] a virtual particle is created and its position and orientation on the surface are selected randomly.
\item[--] the virtual particle is tested for overlaps with any of the other, previously placed, particles. If no overlap is found, it is placed, holding its position and orientation until the end of the simulation. Otherwise, the virtual particle is removed and abandoned.
\end{itemize}
The procedure ends when there is no free space large enough for any other particle to adsorb. The set containing the placed particles is called a saturated random packing. Besides modeling absorption layers, such packings have a large variety of applications including soft matter~\cite{evans1993,torquato2010}, telecommunication~\cite{hastings2005}, information theory~\cite{coffman1998} and mathematics~\cite{zong2014}. 

The scientific history of these packings begins in 1939 when Flory studied reactions between substituents along a long, linear polymer~\cite{flory1939}. In 1958 R{\'e}nyi calculated analytically the mean packing fraction for infinitely large, one-dimensional, saturated random packing, the so-called car-parking problem~\cite{renyi1958}. For higher dimensional packings, the mean packing fraction is known only from numerical simulations. However, numerically generated random packings are always finite, so that, finite size effects can influence its properties. It is especially important, when results of adsorption experiments performed on macroscopic systems are interpreted in terms of numerical modeling. Therefore, it is important to know when finite-size effects can be neglected or not. It is expected that these effects are less important for larger packings than for smaller ones, but on the other hand, larger packings are much harder to obtain because RSA requires then significantly larger number of iterations \cite{ciesla2017}. 

The main goal of this study is to determine how different boundary conditions influences the mean packing fraction and what is the minimal size of saturated random packing that produces the correct value of the mean packing density within a reasonable statistical error. Results of this study can be used to increase effectiveness of RSA modeling, which can be important for searching packing of desired properties, in particular, for finding shapes that gives the highest packing fraction~\cite{ciesla2015shapes,ciesla2016shapes}. The properties of finite random packing with open boudaries was studied previously e.g.~\cite{Adamczyk2007}, but rather in context of particles density fluctuations near boundaries. The periodic boundary conditions seems to be more popular e.g. \cite{feder1980,vigil1989,zhang2013,ciesla2016}, but there are no hints about packing size that is large enough.
\section{\label{sec:model}Model}
We studied RSA of disks on squares with periodic, open, and wall boundary conditions. Each disk had a unit surface area, so its diameter was equal to $d = \sqrt{4/\pi} \approx 1.13$. The surface area of the square boundary $S = L^2$ varied from $6$ to $40000$. To determine  the mean packing fraction $\theta$ for each packing size several independent packings were generated. According to \cite{ciesla2016}, the statistical error of mean packing fraction for periodic boundary conditions is expected to behave as:
\begin{equation}
\sigma (\theta) =  k \sqrt{\frac{\theta}{N_{tot}}} 
\end{equation}
where $N_{tot}$ is a total number of particles in all packings used for determination of $\theta$. The constant $k\approx 0.57$ is for two-dimensional packings. To get similar precision of $\theta$ the number of independent packing generated for each packing size varied form $5\cdot 10^7$ for the smallest packing to $10^4$ for the largest one. This guaranteed that the absolute statistical error did not exceeded $10^{-5}$ for periodic boundary conditions and was at a similar level for other boundary conditions. To generate saturated random packings the improved version of the RSA algorithm described in~\cite{zhang2013} was used. This method significantly reduces simulation time due to excluding regions where there are no possibility of placing another particle. Therefore, the number of RSA trials that ends without adding a particle to the packing is greatly reduced. The idea comes from earlier works concerning RSA on discrete lattices~\cite{nord1991} and deposition of oriented squares on a continuous surface~\cite{brosilow1991}. In~\cite{zhang2013}, the idea was improved by an increase in the precision used to decribe remaining areas, where placing is potentially possible. For spherically symmetric particles each such area will either be filled by a disk or dissapear. When the last area vanishes the packing is saturated. It is worth noting that this algorithm also works for packings of different dimensions.
\section{\label{sec:results}Results}
\begin{figure}[htb]
\centerline{%
\subfigure[]{\includegraphics[width=0.35\columnwidth]{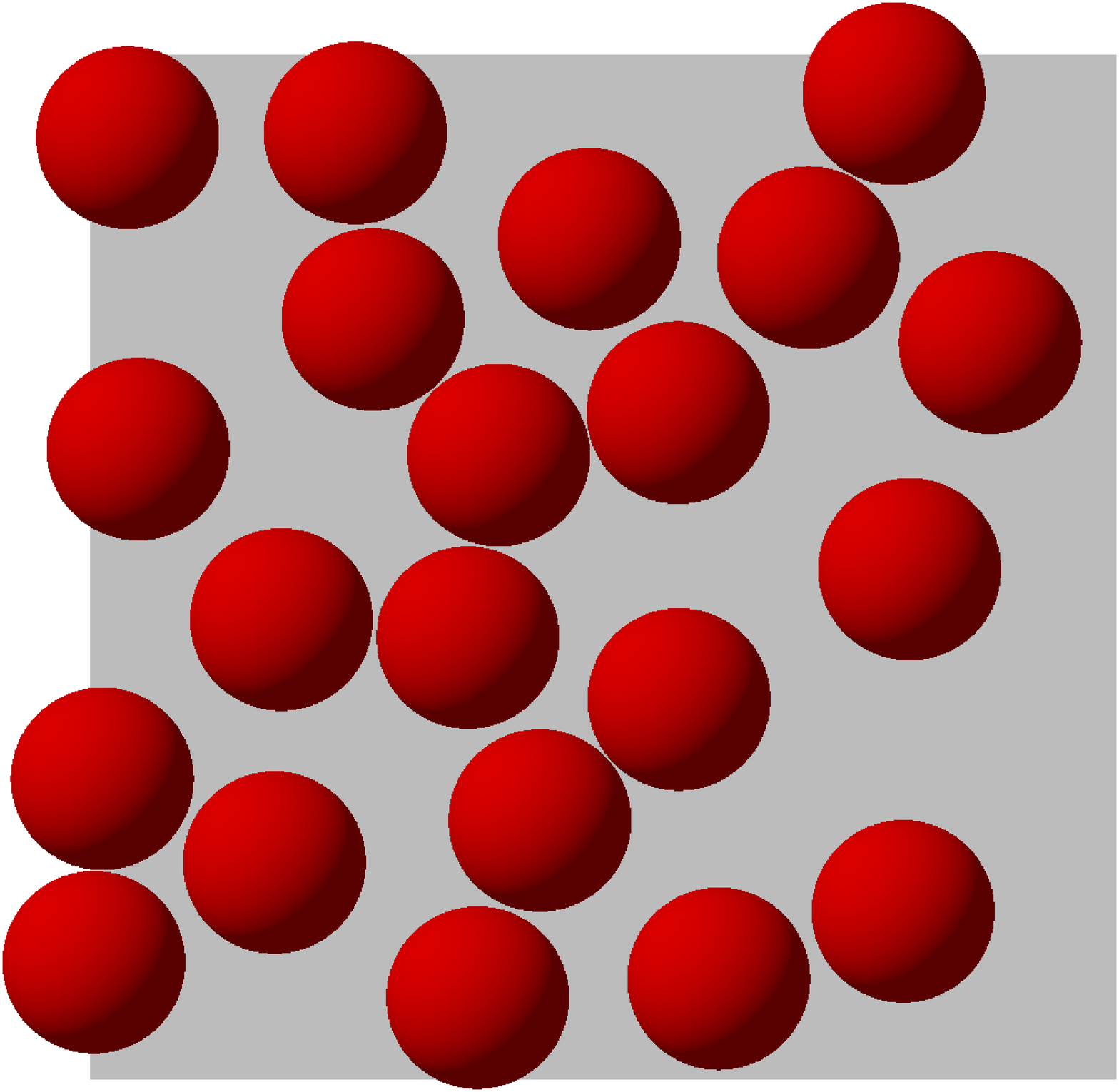}}
\subfigure[]{\includegraphics[width=0.35\columnwidth]{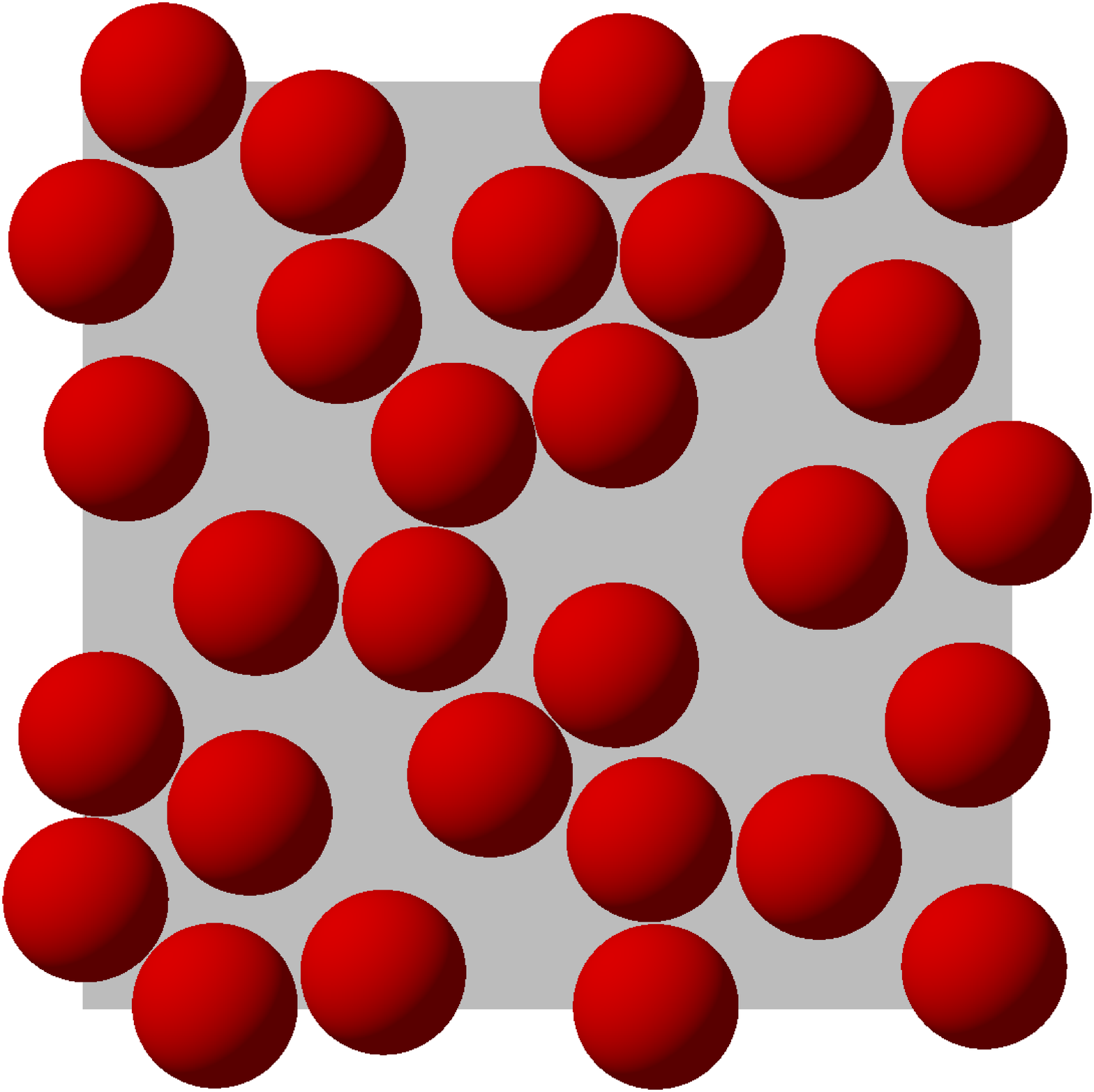}}
\subfigure[]{\includegraphics[width=0.35\columnwidth]{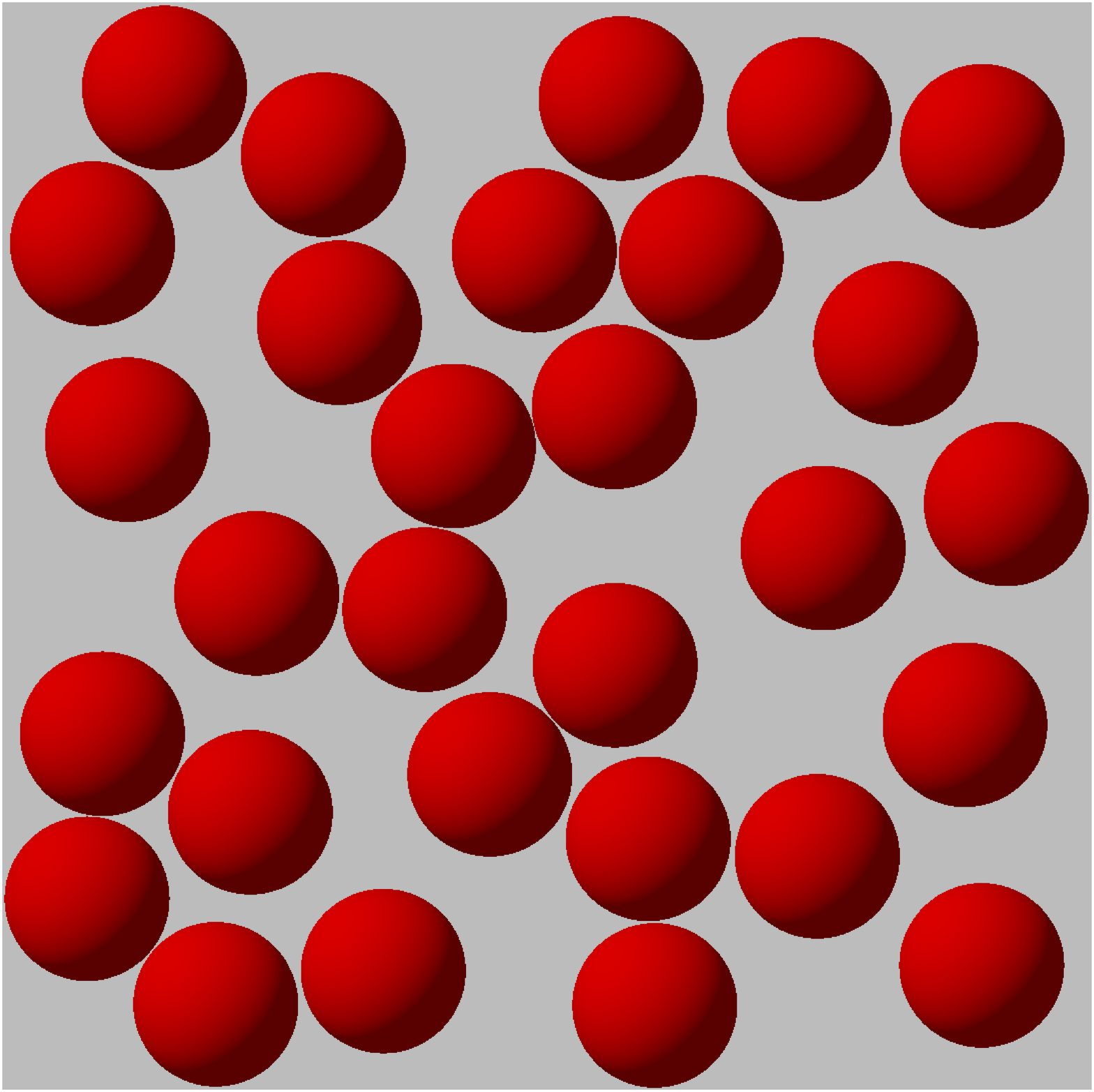}}
}
\caption{Example finite saturated packing using periodic (a), open (b), and wall (c) boundary conditions. The gray square corresponds to the boundary of the packing. Packing sizes are $S=40$ for (a) and (b), and $S=55.546$ for (c).}
\label{fig:packings}
\end{figure}

Fig.\ \ref{fig:packings} presents saturated random packings using periodic (a), open (b), and wall (c) boundary conditions.
Note that disks configurations in packings (b) and (c) are the same. The only difference between them is the packing size. For (b) it is equal to $S=40$ while for (c) $S=55.546$. This difference is a result of condition used to determine if particle is in packing. In the case (b) (open boundary conditions) the particle center has to be within the boundaries, while for (c) (walll boundary conditions) the entire particle has to fit within given square. Thus, the difference between open and wall boundary conditions for the same packing corresponds to a difference of the packing side length equal to particle diameter $d$.
\subsection{Periodic boundary conditions}
\begin{figure}[htb]
\centerline{%
\includegraphics[width=0.6\columnwidth]{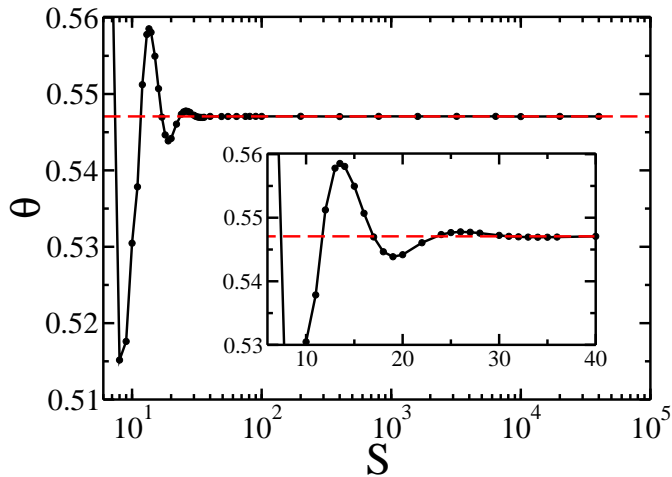}
}
\caption{Dependence of the mean packing fraction on packing size. Dots are data taken from simulations. Statistical errors are approximately $10^{-5}$ for all points and are much less than the size of the dots. The red dashed line corresponds to $\theta=0.547067$. The black solid line is to guide the eye. Inset shows the same data, but focuses on small packing.}
\label{fig:q_s}
\end{figure}

The mean packing fraction measured for the largest packings ($S>100$) with periodic boundary conditions is $\theta = 0.547067 \pm 0.000003$. which agrees with recent values $(0.547074\pm 0.000003$ \cite{zhang2013} and  0.547070 \cite{ciesla2016}). The dependence of the mean packing fraction on packing size is shown in Fig.\ \ref{fig:q_s}.
As expected, for very small packings, finite size effects significantly affects obtained results. What is interesting is that this dependence is not monotonic. The mean packing fraction reaches its maximum around $S=14$. The side length of such square is only $3.3$ times larger than the disk diameter. The next minimum is observed for $S=18$. With the growth of packing size the mean-packing-fraction oscillations dampen quickly and for $S=40$ obtained value of $\theta$ agrees within statistical error with one measured for the largest packings. Note, that $S=40$ corresponds to a square for which side length $L$ is only $5.6$ times larger than the disk diameter $d$.

The reason for the oscillations is presumably that the finite size of the periodic repeat cell allows for a more ordered system compared to an infinite system. The ordering evidently enhances the coverage for some sizes and depresses it for other sizes, because the number of particles that can adsorb is discrete and is sometimes beneficial for increasing the coverage. On the other hand, the density of particles at a given distance is given by the density autocorrelation function: 
\begin{equation}
G(r) = \lim_{dr \to 0} \frac{ \left< N(r, r+dr) \right> }{\theta 2\pi r dr },
\end{equation}
where $\left< N(r, r+dr) \right>$ is the mean number of particles with centers at a distance between $r$ and $r+dr$ from a center of a given disk.
Thus, if the packing size corresponds to the maximum or minimum of this function the mean packing fraction should increase or decrease respectively.
\begin{figure}[htb]
\centerline{%
\includegraphics[width=0.6\columnwidth]{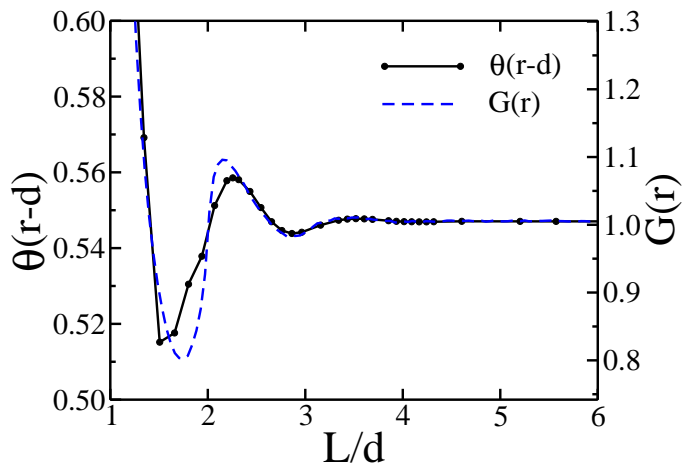}
}
\caption{Comparison between oscillations of the mean packing fraction (black dots and solid line) and density autocorrelation function (blue dashed line). The density autocorrelation function was determined using $100$ independent saturated random packing of a size $S=10^5$ with periodic boundary conditions.}
\label{fig:qcorr_l}
\end{figure}
Indeed, the correspondence is clearly visible in Fig.\ \ref{fig:qcorr_l}. It has been proved analytically that for one-dimensional packings the  oscillations of the density autocorrelation function are damped super-exponentially \cite{bonnier1994}, which explains also why the systematic error originated from finite size effects is negligible for quite small systems.

\begin{figure}[htb]
\centerline{%
\includegraphics[width=0.6\columnwidth]{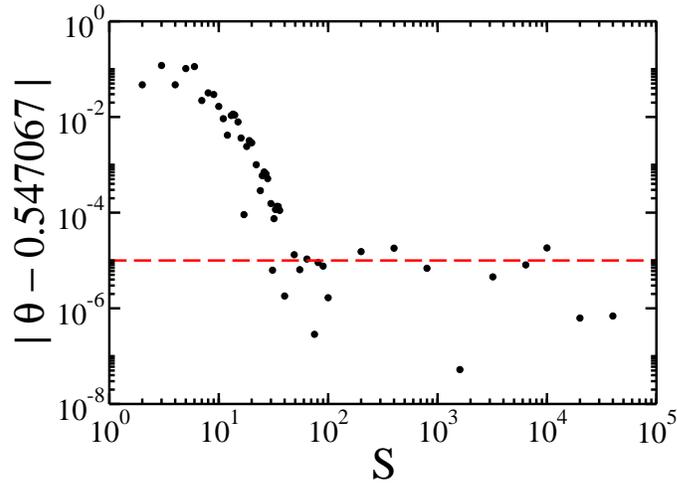}
}
\caption{Dependence of difference between mean packing fraction obtained from simulation and the asymptotic value of $0.547067$ on packing size (black dots). The red dashed line corresponds to statistical error of each value.}
\label{fig:dq_s}
\end{figure}

A direct comparison between the systematic error introduced by finite-size effects and statistical error is shown in Fig.\ \ref{fig:dq_s}.
The plot shows the difference between the best approximation of mean packing fraction obtained for largest packings and its value for each packing of a specific size. As the statistical error of these values is $10^{-5}$ it is clear that even for $S=30$ ($L \approx 4.85 \, d$) the mean packing fraction agrees with its true value within the statistical error. 

\subsection{Open and wall boundary conditions}
\begin{figure}[htb]
\centerline{%
\includegraphics[width=0.6\columnwidth]{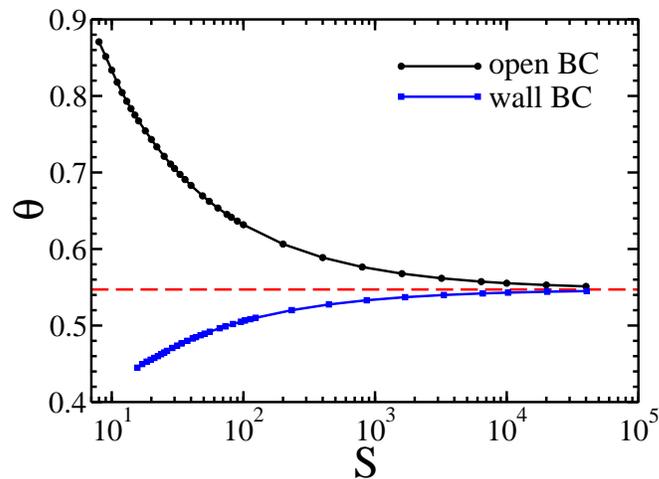}
}
\caption{Dependence of the mean packing fraction on packing size. Black dots and blue squares correspond to free boundary conditions and wall boundary conditions, respectively. Statistical errors are approximately $10^{-5}$ for all points and are much less than the size of the dots. The red dashed line corresponds to $\theta=0.547067$. The blue and black solid lines are to guide the eye.}
\label{fig:qfree_s}
\end{figure}

The mean packing fraction dependence on packing size for open and wall boundary conditions is presented in Fig.\ \ref{fig:qfree_s}. 
As expected, the mean packing fraction is overestimated when using open boundaries and underestimated for wall boundaries but even for the largest packings the difference between obtained packing fractions and $\theta=0.547067$ is two orders of magnitude larger than statistical error.

\begin{figure}[htb]
\centerline{%
\includegraphics[width=0.6\columnwidth]{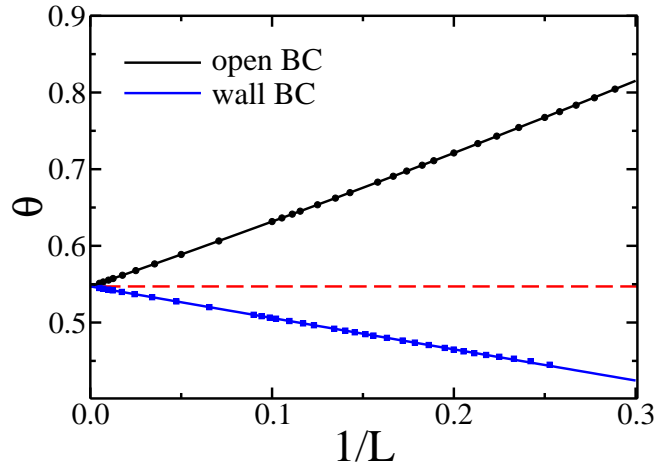}
}
\caption{Dependence of the mean packing fraction on inverse of packing side length $L$ for open and wall boundary conditions. Dots are data and solid lines are fits: $\theta = 0.54707 + 0.82232 \, L^{-1} + 0.23901 \, L^{-2}$ and $\theta = 0.54708 - 0.41285 \, L^{-1} + 0.012711 \, L^{-2}$ for open and wall boundaries, respectively. The red dashed line corresponds to $ \theta = 0.547067$.}
\label{fig:qfree_scaling}
\end{figure}

The typical approach used to estimate real packing fraction is to estimate it using results obtained for finite packings. In such case the dependence of $\theta(1/L)$ can be studied in the limit of $1/L \to 0$. The plot presenting this dependence is shown in Fig.\ \ref{fig:qfree_scaling}.
Although all packing fractions differ significantly from the true value, obtained limiting values from quadratic fits for $L \ge 7$ are within statistical error range for both: open and wall boundaries. Because $\theta = N/L^2$, where $N$ is the number of disks in packing, both these fits can be rewritten in the form:
\begin{equation}
  N(L) = 
  \cases{
  0.54707 \cdot (L + 0.7516)^2 - 0.070 & for open boundaries, \\
  0.54708 \cdot (L - 0.3773)^2 - 0.065 & for wall boundaries. \\
  }
  \label{effective_packing}
\end{equation}
In both cases, we can define the effective packing size $(L_\mathrm{eff})^2$. It is $(L+0.7516)^2$ and $(L-0.3773)^2$ for open and wall boundaries, respectively. Because $0.7515 + 0.3773 \approx d$, the effective size of packing is exactly the same for both cases. Also nearly the same is the negative constant correction which is presumably connected with influence of packing corners. The knowledge about these corrections can help to estimate the mean packing fraction according to the following rule
\begin{equation}
\theta \approx \frac{N + 0.70}{(L_\mathrm{eff})^2}
\label{theta_eff}
\end{equation}
where $L_\mathrm{eff}$ is the effective size of the packing and depends on type of boundary conditions.

\begin{figure}[htb]
\centerline{%
\includegraphics[width=0.6\columnwidth]{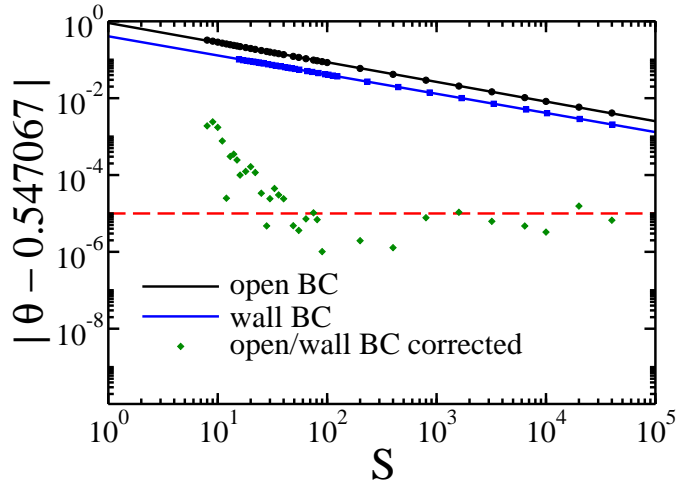}
}
\caption{Dependence of difference between mean packing fraction obtained from simulation and the value of $0.547067$ on packing size. Black dots corresponds to free boundary conditions and blue squares to wall boundary conditions. Solid lines are power fits: $\Delta\theta = 0.907 \, S^{-0.511}$ and $\Delta\theta = 0.408 \, S^{-0.499}$. The green points are the packing fraction calculated using Eq.\ \ref{theta_eff}. The red dashed line corresponds to statistical error of each value.}
\label{fig:dqfree_s}
\end{figure}

The dependence of estimation error on packing sizes for open and wall boundaries are presented in Fig.\ \ref{fig:dqfree_s}. 
It shows that finite size effects decrease with packing size according to a power-law with the exponent $-1/2$. The value of the exponent is a direct consequence of the fact that the area near boundaries scales with packing size as $S^{1/2}$. Thus its relative influence on the whole packing decreases as $S^{1/2} / S = S^{-1/2}$.

According to the power-law the systematic error connected with boundary conditions will reach the level of statistical error for $S \approx 10^9$ or larger. On the other hand, packing fraction estimated using Eq.\ \ref{theta_eff} are almost as good as ones obtained using periodic boundary conditions. Here, the level of statistical error is reached near $S=80$, which corresponds to $L \approx 8d$. However, it should be noted that here we neglected finite precision of $L_\mathrm{eff}$, thus, for accurate estimation of packing fraction from packing with open or wall boundaries the precise values of parameters in Eq.\ (\ref{effective_packing}) is needed.

\section{\label{sec:summary}Summary}
Comparison of periodic, open, and wall boundary conditions shows that the best estimation of the mean packing fraction can be achieved using periodic boundary conditions. In this case, even quite small packing gives reliable results. For a square packing of a side length $5$ times larger than a particle diameter, the systematic error originated in finite size effects is not larger than the statistical error, which in this study was equal to $10^{-5}$. For smaller packings the mean packing fraction oscillates. It seems that character of these oscillations is similar to oscillations of the density autocorrelation function. It can help in designing numerical experiments for packing of different shapes as well as for packing in higher dimensions.

Open and wall boundaries conditions gives results which seems to be  much worse, but there is a systematic way to account finite size effects in calculations by using the effective packing size and constant correction term. It leads                                                                                                                                                                                                                                                                                                                                                                                                                                                                                                                                                                                                                                                                                                                                                                                                                                                                                                                                                                                                                                                                                                                                                                                                                                                                                                                                                                                                                                                                                                                                                                                                                                                                                                                                                                                                                                                                                                                                                                                                                                                                                                                                                                                                                                                                                                                                                                                                                                                                                                                                                                                                                                                                                                                                                                                                                                                                                                                                                                                                                                                                                                                                                                                                                                                                                                                                                                                                                                                                                                                                                                                                                                                                                                                                                                                                                                                                                                                                                                                                                                                                                                                                                                                                                                                                                                                                                                                                                                                                                                                                                                                                                                                                                                                                                                                                                                                                                                                                                                                to results which are almost as good as ones obtained using periodic boundary conditions. Here, the length of a square packing side should be approximately $8$ times larger than particle diameter to reduce systematic error below $10^{-5}$.

Finally it is worth noting that in a typical adsorption experiment, the packing fraction is determined with a precision of only a few percent, which is much lower accuracy than that of the simulations presented here. Therefore, if in numerical modeling one is interested only in predicting the experimental number of adsorbed particles, then RSA of quite small packings will give the proper value.
\section*{Acknowledgements}
This work was supported by Grant No. 2016/23/B/ST3/01145 of the National Science Centre, Poland.
\section*{References}
\bibliographystyle{iopart-num}
\bibliography{main}
\end{document}